\documentclass[twocolumn]{article}

\usepackage{amssymb}
\usepackage{array}
\usepackage{amsmath}
\usepackage{bm}
\usepackage{float}
\usepackage{graphicx}
\usepackage{graphics}
\usepackage{epstopdf}
\usepackage{color}
\usepackage{xfrac}
\usepackage{mathrsfs}
\usepackage{physics}
\usepackage{dsfont}
\usepackage{pdflscape}
\usepackage{lineno}
\usepackage{booktabs}
\usepackage{caption}
\usepackage{authblk}
\usepackage[margin=0.9in]{geometry}
\allowdisplaybreaks

\newcommand*\patchAmsMathEnvironmentForLineno[1]{%
  \expandafter\let\csname old#1\expandafter\endcsname\csname #1\endcsname
  \expandafter\let\csname oldend#1\expandafter\endcsname\csname end#1\endcsname
  \renewenvironment{#1}%
     {\linenomath\csname old#1\endcsname}%
     {\csname oldend#1\endcsname\endlinenomath}}%
\newcommand*\patchBothAmsMathEnvironmentsForLineno[1]{%
  \patchAmsMathEnvironmentForLineno{#1}%
  \patchAmsMathEnvironmentForLineno{#1*}}%
\AtBeginDocument{%
\patchBothAmsMathEnvironmentsForLineno{equation}%
\patchBothAmsMathEnvironmentsForLineno{align}%
\patchBothAmsMathEnvironmentsForLineno{flalign}%
\patchBothAmsMathEnvironmentsForLineno{alignat}%
\patchBothAmsMathEnvironmentsForLineno{gather}%
\patchBothAmsMathEnvironmentsForLineno{multline}%
}

\title{A Coupled Time Domain Random Walk Approach for Transport in Media Characterized by Broadly-distributed Heterogeneity Length Scales}

\author{Tom\'{a}s Aquino\thanks{Corresponding author: tomas.aquino@idaea.csic.es} }
\author{Marco Dentz}
\affil{Spanish National Research Council (IDAEA -- CSIC), 08034 Barcelona, Spain}
\date{}

\providecommand{\e}[1]{\ensuremath{\cdot 10^{#1}}}

\begin{document}

\maketitle

\abstract
We develop a time domain random walk approach for conservative solute transport in heterogeneous media where medium properties vary over a distribution of length scales. The spatial transition lengths are equal to the heterogeneity length scales, and thus determined by medium geometry. We derive analytical expressions for the associated transition times and probabilities in one spatial dimension. This approach determines the coarse-grained solute concentration at the interfaces between regions; we derive a generalized master equation for the evolution of the coarse-grained concentration and reconstruct the fine-scale concentration using the propagator of the subscale transport mechanism. The performance of this approach is demonstrated for diffusion under random retardation in power-law media characterized by heavy-tailed lengthscale and retardation distributions. The coarse representation preserves the correct late-time scaling of concentration variance, and the reconstructed fine-scale concentration is essentially identical to that obtained by direct numerical simulation by random walk particle tracking.





\section{Introduction}

Physical and chemical heterogeneity, which often spans multiple scales, has important consequences for solute transport in natural and engineered media. It is well known that heterogeneity may lead to anomalous (non-Fickian) characteristics, even if the transport mechanism is advective or diffusive at smaller scales~\cite{klages2008anomalous,DCSB2004,havlin1987diffusion,bouchaud1990anomalous,comolli2016non}. Upscaling transport dynamics is essential for understanding, and providing efficient methods for predicting, large-scale solute transport. This is particularly true in view of computational constraints and incomplete information about medium properties.
 
The continuous time random walk (CTRW) framework provides analytical and computational tools for describing transport by considering (conceptual) Lagrangian particles whose movement is characterized by spatial jumps and inter-jump waiting times~\cite{scher1973stochastic,berkowitz2006modeling}. The CTRW as an average transport framework encodes information about the variability in transport dynamics due to subscale heterogeneity through the statistical properties of transition times and distances. Derivation of CTRW-type large-scale descriptions typically requires averaging over an ensemble of medium, or heterogeneity, realizations~\cite{scher1973stochastic,KLSI80.2,bouchaud1990anomalous,BS1997,painter2005upscaling,DCa:2009,comolli2017}. 
Here, we use the term time domain random walk (TDRW) to refer to CTRW approaches which solve transport in single medium representations~\cite{McCarthy:1993,banton1997new,Delay2001,james2001efficient,Painter2008,dentz2012diffusion,noetinger2016}. Transport properties at a given spatial location are fixed, and a particle revisiting a location will sample the same properties. 

\begin{figure}[!htb]
\centering
\includegraphics[width=1\columnwidth]{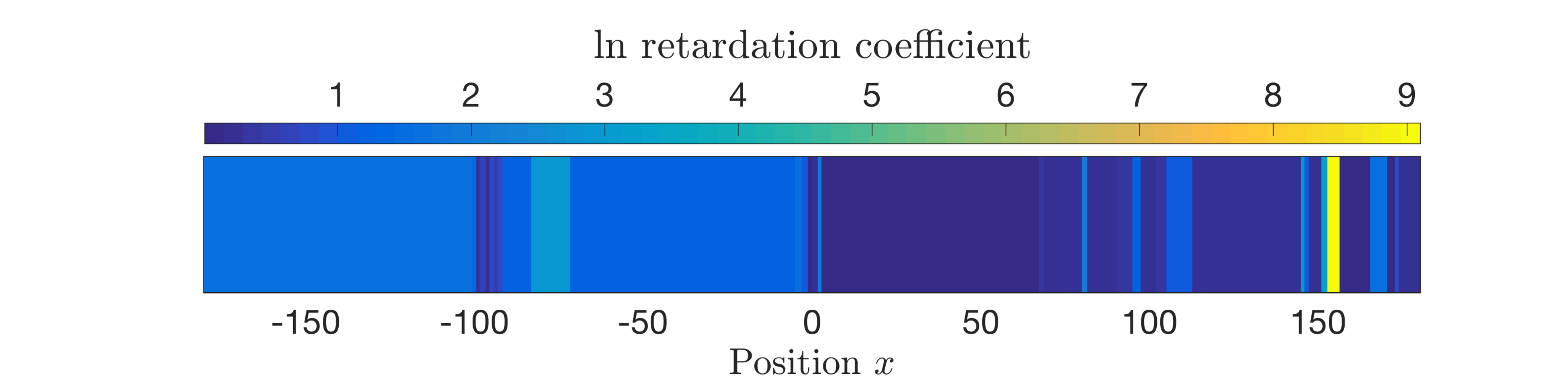}
\caption{Illustration of a one-dimensional medium with power-law-distributed region-dependent retardation coefficient and power-law-distributed region lengths.}
\label{f:medium}
\end{figure}

We consider a one-dimensional medium characterized by a broad distribution of heterogeneity length scales and transport properties, as illustrated in Fig.~\ref{f:medium}. Specifically, we consider spatially variable advection and dispersion as a result of heterogeneous retardation and analyze dispersive trapping and regular dispersion as the small-scale transport mechanisms~\cite{bouchaud1990anomalous}. We construct a TDRW description with spatial transitions over the heterogeneity length scales and derive transition times and transition probabilities that encode the statistical properties of the subscale dynamics. These transition times and probabilities depend on the direction of the transition, resulting in a coupled TDRW model. This is in contrast to models where jump directions are uniformly distributed~\cite{russian2017self,massignan2014nonergodic,dentz2016self} and/or transition times are assumed to be independent of the jump direction~\cite{dentz2012diffusion}.

TDRW descriptions based on finite-volume discretizations of the
advection--dispersion equation (ADE) do not resolve the particle
position within a pixel or voxel. This gives rise to numerical
dispersion, which can be addressed by refining the
discretization~\cite{russian2016}. For transport in a medium whose
properties are distributed on a hierarchy of macroscopic length
scales, the subscale process needs to be accounted for in order to
accurately represent the particle position and thus the concentration
distribution. We derive a procedure for reconstructing the fine-scale
concentration from the coarse-grained particle distribution obtained
from the TDRW. In this sense, the resulting model represents a computationally efficient, particle-based, hybrid approach, in that it combines fast coarse-scale simulations with an efficient local reconstruction procedure.


\section{Transport models}

Solute concentration $c$ for diffusive transport through a one-dimensional medium with trapping characterized by a position-dependent retardation coefficient $\theta$ obeys the Fokker--Planck equation~\cite{bouchaud1990anomalous,risken1996fokker}
\begin{align}
\label{eq::trapping}
	\frac{\partial c(x,t)}{\partial t} = - \frac{\partial}{\partial x}[v(x)c(x,t)] + \frac{\partial^2}{\partial x^2}[D(x) c(x,t)] ,
\end{align}
where $v(x) = v_0/\theta(x)$ is the transport velocity and $D(x) = \kappa/\theta(x)$, with $\kappa$ the constant (molecular) diffusion coefficient and $v_0$ the constant flow velocity. Note that this is not a regular dispersion equation (except in the case of homogeneous retardation), because $\partial^2[D(x)c(x,t)]/\partial x^2 \neq \partial^2[D(x)\partial c(x,t)/\partial x]/\partial x^2$. Subsequently, we will call it the trapping equation. Equivalently, transport may be described by the Langevin equation for particle trajectories $X$,
\begin{equation}
\label{eq::ptrw}
	d X(t) = v[X(t)] d t + \sqrt{2D[X(t)]d t} \, \xi(t) ,
\end{equation}
where, for each time $t$, $\xi(t)$ is an independent Gaussian random variable with mean zero and unit variance. This equation is to be interpreted in the It\=o sense~\cite{van1992stochastic}, and applies with independent $\xi$ to each particle. It forms the basis for particle tracking random walk (PTRW) simulations, which we employ below to verify our results for the upscaled TDRW model. Concentration corresponds to the probability density function (PDF) of Lagrangian particle positions scaled by the total mass. Throughout, we normalize concentrations to unit mass, so that the spatial integral of concentration is equal to $1$ at all times. The Fokker--Planck equation for the PDF of particle position corresponding to~\eqref{eq::ptrw} coincides with the trapping equation~\cite{bouchaud1990anomalous}.

For transport under spatially variable advection and regular dispersion (as opposed to trapping), the Fokker--Planck equation is the ADE,
\begin{equation}
\label{eq::ADE}
\frac{\partial c(x,t)}{\partial t} = - \frac{\partial}{\partial
  x}[v(x) c(x,t)] + \frac{\partial}{\partial x}\left[D(x)
\frac{\partial  c(x,t)}{\partial x}\right].
\end{equation}
The corresponding Langevin equation is given by~\cite{noetinger2016},
\begin{align}
\notag
d X(t) &= \left(v[X(t)] + \frac{dD[X(t)]}{dx}\right) \, dt\\
&\quad+ \sqrt{2D[X(t)] \, dt} \, \xi(t) .
\end{align}
The dynamics for spatially discontinuous dispersion can be integrated numerically using Eq.~\eqref{eq::ptrw} along with a predictor--corrector method~\cite{labolle2000diffusion}.

\section{Coarse graining}

We coarse-grain transport by considering a time domain random walk (TDRW) in a medium composed of segments characterized by a length and a constant retardation coefficient. We take particles to start at a given node between two segments; at each step $n$, particles wait for a random time $T_n$ and then jump a length $L_n$ to an adjacent node.
We thus define our TDRW by the Lagrangian equations
\begin{subequations}
\label{eq::tdrw}
\begin{align}
	X_{n+1} &= X_n + L_n(X_n) ,\\
	T_{n+1} &= T_n + \tau_n(X_n,L_n) .
\end{align}
\end{subequations}
For $x$ the position of a node, let $\ell_+(x)$ and $\ell_-(x)$ be the lengths of the segments to its immediate right and left, respectively. Denote the probabilities of a jump to the right or left as $p_\pm(x)$. The jump length $L_n(x)$ is characterized by the probabilities $P\{L_n(x) = \pm \ell_\pm(x)\} = p_\pm(x)$. The transition times are given by $\tau_n(x,l) = \Theta(l) \tau_{n,+}(x) + \Theta(-l) \tau_{n,-}(x)$, where $\Theta$ is the Heaviside step function. We denote the PDFs of the travel times $\tau_{n,\pm}(x)$ given the direction of transition as $\psi_\pm(\cdot;x)$. We write also $D_\pm(x) = \kappa / \theta_\pm(x)$, where $\theta_\pm(x)$ are the retardation coefficients of the segments to the right and left of the node, and $v_\pm(x) = v_0/\theta_\pm(x)$ for the corresponding velocities.

\subsection{First arrival times in the unit cell}
\label{s::fpt_unit_cell}

\begin{figure}[!htb]
\centering
\includegraphics[width=1\columnwidth]{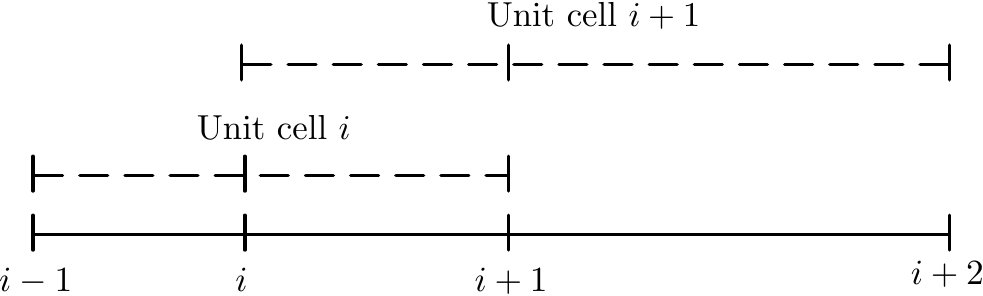}
\caption{Unit cells associated with nodes $i$ and $i+1$.}
\label{f:unit_cell}
\end{figure}

We define a unit cell as composed of a central node and the two adjacent segments, as illustrated in Fig.~\ref{f:unit_cell}.
The transition probabilities $p_\pm$ represent the probabilities of a transported particle starting from the central node to first reach the node to the right or the node to the left. The $\psi_\pm$ represent the PDFs of the first arrival time to the corresponding node, given that node is reached first. Since the TDRW defined by Eq.~\eqref{eq::tdrw} is Markovian in the transition $n$ (no memory of previous transitions) and only allows transitions to adjacent nodes, this fully characterizes the system.

In order to find the first arrival time PDFs, we solve a Green function problem, for Eq.~\eqref{eq::trapping} for the trapping problem and Eq.~\eqref{eq::ADE} for the dispersion problem, with absorbing boundary conditions at the outer edges and a pulse initial condition of unit mass at $x=0$. We choose a coordinate system such that $x=0$ corresponds to the central node, and the edges are located at $-\ell_-<0$ and $\ell_+>0$. The boundary and initial conditions for the Green function (i.e., concentration propagator) $g$ are then
\begin{equation}
	g(-\ell_-,t) = g(\ell_+,t) = 0, \quad g(x,0) = \delta(x),
\end{equation}
where $\delta(\cdot)$ is the Dirac delta. 
Solutions can be written in the form
\begin{equation}
\label{eq::sol_piecewise}
	g(x,t) = g_-(x,t) \Theta(-x) + g_+(x,t) \Theta(x).
\end{equation}
The continuity condition for $g(x,t)$ at the interface at $x = 0$ is obtained from the requirement that concentration be integrable anywhere in the unit cell. This implies for the trapping problem that 
\begin{align}
\label{eq:conttrap}
D_- g_-(0,t)=D_+ g_+(0,t).
\end{align}
For the dispersion problem, we obtain the well-known condition that the concentration be continuous,
\begin{align}
\label{eq:contdisp}
g_-(0,t) = g_+(0,t). 
\end{align}

Details on the calculations of the Green function are given
in~\ref{a::unit_cell}. Once the Green function is known, the fluxes
through the cell boundaries determine the first arrival time densities,
\begin{align}
\phi_\pm(t) = D_\pm \bigg|\frac{\partial g_\pm(x,t)}{\partial x}\bigg|_{x=\pm \ell_\pm}.
\end{align}
Note that $\phi_{\pm}(t) \, dt$ denotes the joint probability of the particle arriving at the right/left cell boundary with an arrival time in $[t,t+dt[$.  
The PDF of residence times in the unit cell is given by 
\begin{align}
\label{psi}
\psi(t) = \phi_{+}(t) + \phi_{-}(t).
\end{align}
The integral of $g(x,t)$ over the unit cell is equal to the probability that a particle has not left the cell by time $t$, i.e., that its residence time is larger than $t$,
\begin{align}
\label{eq:cpsi}
\int\limits_{\Omega} dx \, g(x,t) = \int\limits_t^\infty dt' \, \psi(t'),
\end{align}
where $\Omega = [-\ell_{-}, \ell_{+}]$.
Thus, $g(x,t) \, dx$ is the joint probability that the particle position is in $[x,x+dx]$ at time $t$ and that the particle is still in the cell. 

The probabilities of arriving at the right or left cell boundary first are given by 
\begin{align}
p_{\pm} = \int\limits_0^\infty dt \, \phi_{\pm}(t),
\end{align}
and the PDFs of arrival times at the right or left boundary are given by 
\begin{align}
\psi_{\pm}(t) = \frac{\phi_{\pm}(t)}{p_{\pm}}. 
\end{align}
%

\begin{table*}[tb]
\centering
\caption{Unit cell quantities.}
{\renewcommand{\arraystretch}{1.5}
\begin{tabular}[c]{| c | c | c |}
\hline
& Trapping & Dispersion \\
\hline
$\tilde g_\pm(x,\lambda)$ &
$\sqrt{\frac{D_\mp}{D_\pm}}A(\lambda)\sinh[(\ell_\pm\mp x)v^*_\pm\alpha_\pm(\lambda)]\csch[v^*_\pm\alpha_\pm(\lambda)]$ &
$A(\lambda)\sinh[(\ell_\pm\mp x) v^*_\pm\alpha_\pm(\lambda)]\csch[v^*_\pm\alpha_\pm(\lambda)]$ \\
\hline
$A(\lambda)$ &
$\left[\sum\limits_{\sigma\in\{+,-\}} \!\!\! v_\sigma\sqrt\frac{D_{-\sigma}}{D_\sigma} \left\{ \alpha_\sigma(\lambda) \coth[v^*_\sigma\alpha_\sigma(\lambda)] + [\sigma 1] \right\}\right]^{-1}$ &
$\left[\sum\limits_{\sigma\in\{+,-\}} \!\!\! v_\sigma \left\{ \alpha_\sigma(\lambda) \coth[v^*_\sigma\alpha_\sigma(\lambda)]+[\sigma1]\right\}\right]^{-1}$ \\
\hline
$\tilde\phi_\pm(\lambda)$ &
$\sqrt{\frac{D_\mp}{D_\pm}}v^*_\pm\alpha_\pm(\lambda)A(\lambda)\csch[v^*_\pm\alpha_\pm(\lambda)]$ &
$v^*_\pm\alpha_\pm(\lambda)A(\lambda)\csch[v^*_\pm\alpha_\pm(\lambda)]$ \\
\hline
$p_\pm$ &
$\frac{\sqrt\frac{D_\mp}{D_\pm}|v_\pm|e^{\pm v^*_\pm}\csch(|v^*_\pm|)}{\sum\limits_{\sigma\in\{+,-\}} \!\!\! \sqrt\frac{D_{-\sigma}}{D_\sigma} |v_\sigma| e^{\sigma v^*_\sigma} \csch(|v^*_\sigma|)}$ &
$\frac{|v_\pm|e^{\pm v^*_\pm}\csch(|v^*_\pm|)}{\sum\limits_{\sigma\in\{+,-\}} \!\!\! \!\!\!|v_\sigma| e^{\sigma v^*_\sigma} \csch(|v^*_\sigma|)}$ \\
\hline
&\multicolumn{2}{|c|}{No advection}\\
\hline
$\tilde g_\pm(x,\lambda)$ &
$\sqrt{\frac{D_\mp}{D_\pm}}A(\lambda)\sinh[\lambda^*_\pm \mp \lambda^*(x)]\csch(\lambda^*_\pm)$ &
$A(\lambda)\sinh[\lambda^*_\pm \mp \lambda^*(x)]\csch(\lambda^*_\pm)$ \\
\hline
$A(\lambda)$ &
$\sum\limits_{\sigma\in\{+,-\}} \!\!\! \sqrt{D_{-\sigma} \lambda}\coth(\lambda^*_\sigma)$ &
$\sum\limits_{\sigma\in\{+,-\}} \!\!\! \sqrt{D_\sigma \lambda}\coth(\lambda^*_\sigma)$ \\
\hline
$\tilde\phi_\pm(\lambda)$ &
$\sqrt{D_\mp \lambda}A(\lambda)\csch(\lambda^*_\pm)$ &
$\sqrt{D_\pm \lambda}A(\lambda)\csch(\lambda^*_\pm)$ \\
\hline
$p_\pm$ &
$\frac{\ell_\mp}{\ell_++\ell_-}$ &
$\frac{D_\pm \ell_\mp}{D_-\ell_++D_+\ell_-}$ \\
\hline
\end{tabular}
}
\caption*{Transition probabilities $p_\pm$, Laplace-transformed (denoted by a tilde) first arrival densities $\tilde \phi_\pm$, and concentration propagators $\tilde g$ in the unit cell for trapping and regular dispersion.
We use the notations: $\lambda^*(x) = x^2 \lambda / D_\pm$, $\lambda^*_\pm = \lambda^*(\ell_\pm)$, $v^*_\pm = \ell_\pm v_\pm / (2D_\pm)$, and $\alpha_\pm(\lambda) = \operatorname{sgn}(v_\pm)\sqrt{1+4D_\pm \lambda/v_\pm^2}$, where $\operatorname{sgn}(v)$ denotes the sign of $v$. \label{t::unit_cell}}
\end{table*}

\begin{figure*}[!htb]
\centering
\includegraphics[width=1\textwidth]{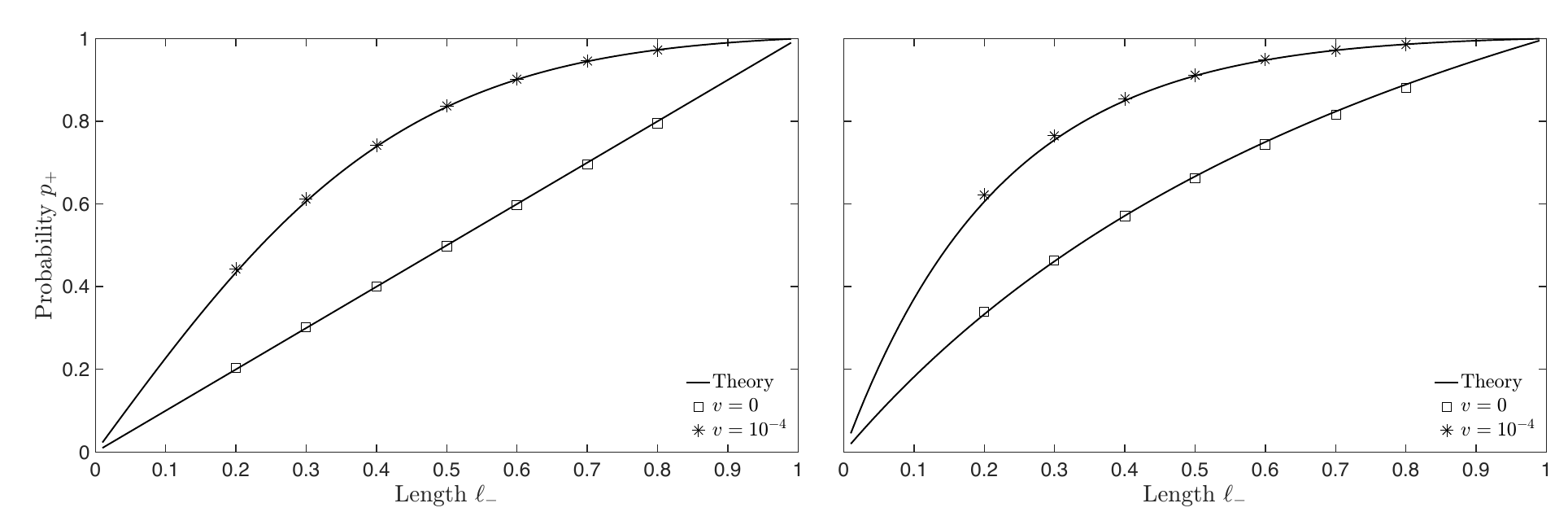}
\caption{Probability $p_+$ of exiting the unit cell through the right boundary as a function of left segment length $\ell_- = 1 - \ell_+$. Symbols are computed from PTRW simulations. Left: trapping, Right: regular dispersion.}
\label{fi::prob}
\end{figure*}

\begin{figure*}[!htb]
\centering
\includegraphics[width=1\textwidth]{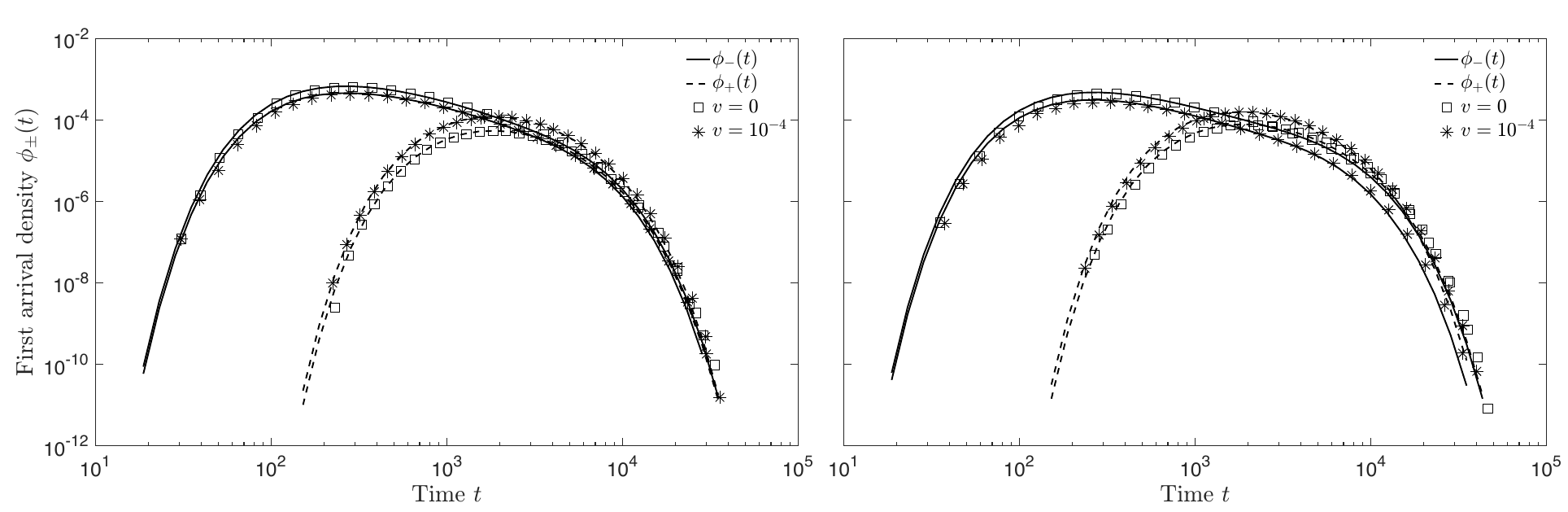}
\caption{First arrival time densities $\phi_\pm$ to the right and left in the unit cell with $\ell_- = 1 - \ell_+ = 0.2$. Symbols are computed from PTRW simulations. Solid and dashed lines ($\phi_-$ and $\phi_+$ respectively) are obtained from numerical inversion of the theoretical results for the Laplace transforms. Left: trapping, Right: regular dispersion.}
\label{fi::fpt}
\end{figure*}

Results are summarized in Table~\ref{t::unit_cell}. We verified the analytical results for the first arrival times and transition probabilities against numerical simulations. The Langevin formulation, Eq.~\eqref{eq::ptrw} (with a predictor--corrector method for regular dispersion~\cite{labolle2000diffusion}), was used to find the exiting boundary and the corresponding first arrival time for $10^7$ trajectories starting at $x = 0$ in a unit cell with $\ell_- + \ell_+ = 1$, using a time step $\Delta t = 1$. Results for the exiting probability $p_+$ and first arrival time densities $\phi_\pm$, for $D_- = 2.5\e{-5}$ and $D_+ = 5\e{-5}$, with $v_+=v_-=v$, for $v=0$ and $v=10^{-4}$, are shown in Figs.~\ref{fi::prob} and~\ref{fi::fpt}.

\subsection{Spatial and temporal transitions in the TDRW} 
\label{s::tdrw_fpt}

We now turn back to the TDRW description~\eqref{eq::tdrw}. Consider the particle position $X_n$ after $n$ TDRW steps to correspond to the $i$th node and thus the $i$th unit cell in the medium. The quantities referring to unit cell $i$ in the following are marked by a subscript $i$. As outlined in the previous Section, the joint probabilities to make a transition from node $i$ to the left or right after a transition time in $[t,t+dt[$ are given by $\phi_{-,i}(t) \, dt$ and $\phi_{+,i}(t) \, dt$, respectively. Denoting the joint probability that a transition from node $j$ to node $i$ occurs after a waiting time in $[t,t+dt[$ by $\psi_{ij}(t) \, dt$, we can thus write
\begin{align}
\psi_{ij}(t) = \delta_{j,i-1} \phi_{+,j}(t) + \delta_{j,i+1} \phi_{-,j}(t),
\end{align}
where $\delta_{\cdot,\cdot}$ is the Kronecker delta. Note that only transitions to nearest neighbors are allowed. The PDF of waiting times at node $i$ can then be written as
\begin{align}
\psi_i(t) = \sum_{[ij]} \psi_{ji}(t),
\end{align}
where $\sum_{[ij]}$ denotes summation over the nearest neighbors $j$ of node $i$. The TDRW algorithm determines first the direction of the transition from the probabilities $p_{\pm,i}$, and the transition lengths are accordingly given by $\ell_{\pm,i}$, respectively. Once the direction of the transition is determined, the waiting time is drawn from the corresponding conditional waiting time PDF $\psi_{\pm,i}$.

In the following, we discuss the transition probabilities and transition time distributions and focus for illustration on the case $v(x) \equiv 0$.
For a more detailed discussion of a TDRW approach dealing with purely advective transport, we refer the interested reader to~\cite{comolli2016non}. In the following, we omit the subscript dependency on the node for compactness of notation.
\subsubsection{Transition probabilities}
The transition probabilities $p_{\pm}$ for the trapping model are given by 
\begin{align}
p_{\pm} = \frac{\ell_{\mp}}{\ell_{+} + \ell_{-}}. 
\end{align}
They are independent of the retardation coefficients
and depend only on the geometry of the unit cell. The probability of a transition to the nearest node is higher. For equal lengths, the
transition probabilities are equal. Note that the steady state
concentration in the unit cell, which is obtained for no-flux conditions at the boundaries, satisfies
\begin{align}
c^\infty_{\pm} = \frac{D_\mp}{D_{\pm}} = \frac{\theta_{\pm}}{\theta_{\mp}}.
\end{align}
This can be directly obtained from the continuity condition~\eqref{eq:conttrap} and is independent of the cell geometry.  
This means particles accumulate in the region of low dispersion, or equivalently of high retardation. 

For the dispersion problem this different. The transition probabilities depend both on the geometry of the unit cell and the dispersion coefficients, 
\begin{align}
\label{pdisp}
p_{\pm} = \frac{D_\pm \ell_\mp}{D_-\ell_++D_+\ell_-}. 
\end{align}
For equal lengths, the probability of a transition in the direction of the higher dispersion coefficient is higher. The steady state concentration in this case obeys
\begin{align}
c^\infty_+ = c^\infty_-,
\end{align}
which is again directly obtained from the continuity condition~\eqref{eq:contdisp}. Equation~\eqref{pdisp} expresses the fact that the transition probabilities must be asymmetric in order to guarantee equidistribution, because the residence times on the left and right side of the unit cell are different and depend inversely on the dispersion coefficient.

\subsubsection{Transition time distributions}

As discussed in Section~\ref{s::tdrw_fpt}, we distinguish the
conditional first arrival time PDFs $\psi_{\pm}$, which
correspond to the times to first arrive at the right or left
boundary, respectively, and the residence time PDF~\eqref{psi} in the unit cell, which
can also be written as
\begin{align}
\psi(t) = p_+ \psi_+(t) + p_- \psi_-(t). 
\end{align}
The average conditional transition times are defined by
\begin{align}
\mu_{\pm} = \int\limits_0^\infty dt \, t \psi_{\pm}(t). 
\end{align}
They can be obtained by differentiation from the results given in
Table~\ref{t::unit_cell} as
\begin{align}
\mu_\pm = \left.\frac{d \tilde \psi_\pm(\lambda)}{d \lambda}
  \right|_{\lambda = 0}. 
\end{align}
Here and throughout, we denote the Laplace transform of a function with respect to the time variable by a tilde. We obtain the explicit expressions
\begin{align}
\mu_\pm = \frac{\ell_\pm^2}{2D_{\mathrm{eff},\pm}}, && D_{\mathrm{eff},\pm}^{-1} = \frac{1}{3D_\pm} + \frac{2\ell_\mp}{3\ell_\pm D_{\mathrm{eff}}},
\end{align}
where the effective dispersion coefficients $D_{\mathrm{eff}}$ for the trapping and dispersion problems are given respectively by
\begin{align}
	D_{\mathrm{eff}} = \frac{\kappa}{\theta_A}, && 	D_{\mathrm{eff}} = D_A,
\end{align}
with $\theta_A$ and $D_A$ the spatial arithmetic mean retardation and dispersion coefficients in the unit cell.
The first arrival time PDFs and the mean arrival times in either direction depend on the properties to the immediate left and right of a node because particles may sample both regions before they arrive at one of the two boundaries. The average waiting time is given by 
\begin{equation}
	\mu =  p_+\mu_+ + p_-\mu_- = \frac{\ell_- \ell_+}{2D_{\mathrm{eff}}}.
\end{equation}
The variances of the conditional arrival times are obtained from the Laplace transforms $\tilde \psi_\pm$ through 
\begin{align}
\sigma_\pm^2 = \left.\frac{d^2 \ln[\tilde \psi_\pm(\lambda)]}{d \lambda^2}\right|_{\lambda = 0}. 
\end{align}
For the trapping problem, we have
\begin{align}
\notag
	\sigma_\pm^2 &= \frac{\ell_-^2 \ell_+^2}{6}\Bigg[ \frac{2}{3D_{\mathrm{eff}}^2} + \frac{\ell_\pm^2}{15 \ell_\mp^2 D_\pm^2}\\
	&\quad+ \frac{4}{15}\frac{D_-^2\ell_+^3+D_+^2\ell_-^3}{D_-^2D_+^2\ell_-\ell_+(\ell_++\ell_-)} \Bigg],
\end{align}
and for regular dispersion
\begin{align}
\notag
	\sigma_\pm^2 &= \frac{\ell_-^2 \ell_+^2}{6}\Bigg[ \frac{2}{3D_{\mathrm{eff}}^2} + \frac{\ell_\pm^2}{15 \ell_\mp^2 D_\pm^2}\\
	&\quad+ \frac{4}{15}\frac{D_-\ell_+^3+D_+\ell_-^3}{D_-D_+ D_{\mathrm{eff}}\ell_-\ell_+(\ell_++\ell_-)} \Bigg].
\end{align}

\begin{figure*}[!htb]
\centering
\includegraphics[width=1\textwidth]{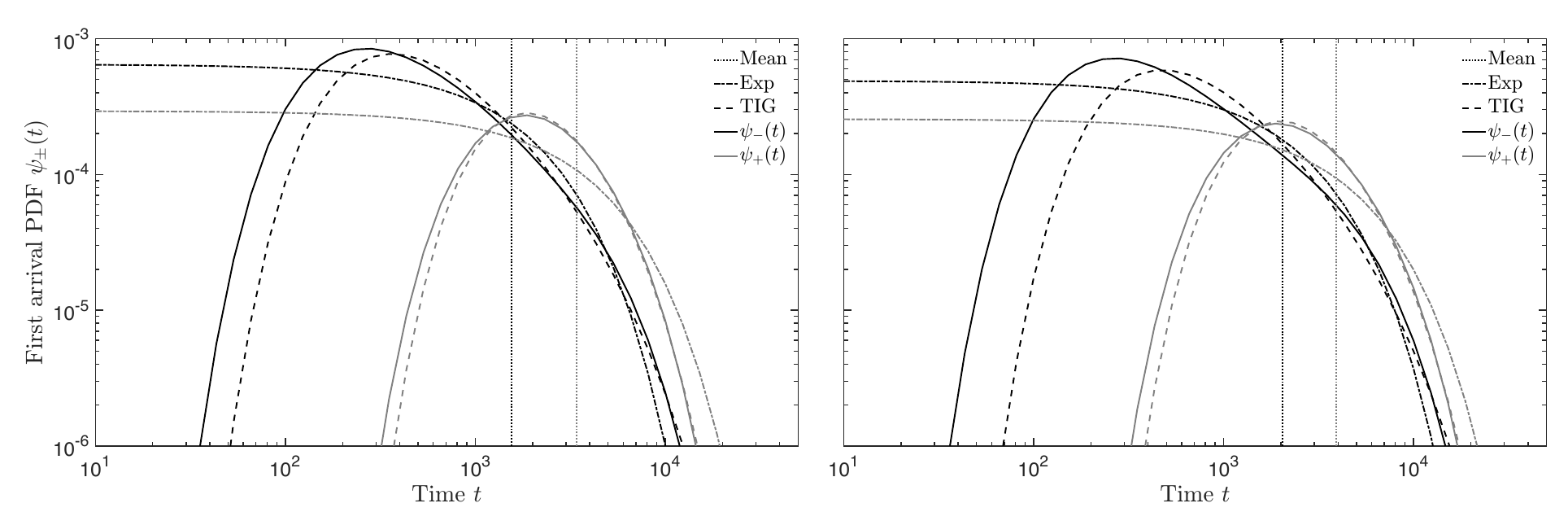}
\caption{The three approximations of the conditional first arrival time PDFs in the unit cell discussed in the text, with $v=0$, $D_-=2.5\e{-5}$,  $D_+5=\e{-5}$, and $\ell_-=1-\ell_+=0.2$. The solid lines are obtained by numerically inverse Laplace transforming the analytical solutions.}
\label{f:fpt_approx}
\end{figure*}

In the absence of a closed form solution for the conditional arrival time PDFs $\psi_\pm$ calculated above in Laplace space, numerically integrating the full model requires numerically inverting a large number of Laplace transforms. We thus consider three different approximations for $\psi_\pm(t)$. First, we approximate the conditional arrival times by their mean, so that
\begin{equation}
\label{eq::fpt_mean}
	\psi_\pm(t) = \delta(t-\mu_\pm).
\end{equation}
Second, we approximate the $\psi_\pm$ by exponential PDFs as
\begin{equation}
\label{eq::fpt_exp}
	\psi_\pm(t) =  \frac{e^{-t/\mu_\pm}}{\mu_\pm}.
\end{equation}
This simple approximation retains the correct average, while the variance is different from the true variance. 
Finally, we wish to find an approximation with an appropriate functional form which matches the mean and variance of the exact PDFs. We take into account that the $\psi_\pm$ are highly skewed, with an intermediate diffusive $t^{-3/2}$ scaling and an exponential cutoff, as shown in Fig.~\ref{fi::fpt}. Thus, we approximate them using tempered L\'evy-stable PDFs of order $1/2$, which are truncated inverse Gaussians (TIGs)~\cite{meerschaert2012stochastic},
\begin{equation}
\label{eq::fpt_tig}
	\psi_\pm(t) = \sqrt\frac{\mu_\pm^3}{2\pi\sigma_\pm^2t^3}e^{-\frac{\mu_\pm}{2\sigma_\pm^2t}(t-\mu_\pm)^2}.
\end{equation}
The two independent parameters are fixed so as to match the analytical mean and variance. A comparison between the three approximations and the exact arrival time PDFs is shown in Fig.~\ref{f:fpt_approx}.

\section{Reconstruction of fine-scale concentration}

In this Section, we develop a procedure for the reconstruction of the fine-scale concentration within each unit cell. To this end, we first obtain a generalized master equation~\cite{KMS73} for the evolution of the concentration corresponding to the TDRW model defined by Eq.~\eqref{eq::tdrw}. We then obtain the governing equation for the fine-scale concentration and discuss the reconstruction procedure.

\subsection{Coarse-grained concentration}

The TDRW~\eqref{eq::tdrw} describes transitions between the nodes that separate regions of a medium characterized by different transport properties. The coarse-grained mass is thus concentrated at the inter-region nodes. The full transition time PDF at node $i$ is $\psi_i$. Thus, the probability $P_i(t)$ for a particle to be at node $i$ at time $t$ is given by~\cite{berkowitz2006modeling, dentz2012diffusion} 
\begin{align}
\label{eq:P}
P_i(t) = \int\limits_0^t dt' R_i(t') \int\limits_{t-t'}^\infty dt'' \psi_i(t''),
\end{align}
where $R_i(t')$ is the probability per time that a particle has just arrived at node $i$ at time $t'$, which multiplies the probability that the transition away from node $i$ takes longer than $t - t'$. 
The $R_i$ satisfy the Kolmogorov-type equation
\begin{align}
\label{eq:R}
R_i(t) = R_0(t) + \sum_{[ij]}\int\limits_0^t dt' R_j(t') \psi_{ij}(t-t'),
\end{align}
where $R_0(t) = \delta_{i,0}\delta(t)$ is determined by the initial condition. Equations~\eqref{eq:P} and~\eqref{eq:R} can be combined into the generalized master equation
\begin{align}
\notag
\frac{d P_i(t)}{dt} &= \sum_{[ij]} \int\limits_0^t dt' \bigg[\Phi_{ij}(t-t') P_{j}(t')\\
&\quad - \Phi_{ji}(t-t') P_i(t') \bigg],
\label{eq::meq_waiting}
\end{align}
where the memory kernel $\Phi_{ij}$ is defined in Laplace space by
\begin{align}
\tilde \Phi_{ij}(\lambda) = \frac{\lambda \tilde \psi_{ij}(\lambda)}{1 - \tilde \psi_j(\lambda)}. 
\end{align}
%
\subsection{Fine-scale concentration}
\label{s::interp}

The TDRW description~\eqref{eq::tdrw} fully represents the transition time and transition probability dynamics of the subscale transport mechanism. However, it interprets particles as staying at a node position before making a jump and thus does not resolve the particle positions inside each unit cell. The reconstruction of the fine-scale concentration $c_i$ associated with node $i$ is given by
\begin{align}
\label{eq:pi}
c_i(x,t) = \int\limits_0^t dt' R_i(t') g_i(x - x_i,t-t'),
\end{align}
where $g_i$ is the concentration propagator in the $i$th unit cell, see Section~\ref{s::fpt_unit_cell}. Thus, $g_i(x,t) \, dx$ denotes the joint probability that a particle is still in the unit cell $i$ at time $t$ and at a position in $[x,x+dx[$ relative to $x_i$, see also~\eqref{eq:cpsi}.
It is clear that Eq.~\eqref{eq:P} is obtained from~\eqref{eq:pi} by integration over the unit cell. Using~\eqref{eq:P}, we can express the fine-scale concentration $c_i$ in terms of the coarse-grained probability $P_i$. We obtain the explicit Laplace-space expression
\begin{equation}
\label{eq::interp}
	\tilde c_i(x,\lambda) = \frac{\lambda \tilde g_i(x-x_i,\lambda)}{1 - \tilde \psi_i(\lambda)} \tilde P_i(\lambda), 
\end{equation}
where we have used the Laplace transform of~\eqref{eq:P}. The concentration $c(x,t)$ is given by the superposition of the contributions $c_i(x,t)$ due to all nodes,
\begin{align}
\label{eq:c_fine}
c(x,t) = \sum_i c_i(x,t). 
\end{align}
Note that the concentration at any given position $x$ is determined by the contributions of the two nearest nodes.

\subsection{Numerical implementation}

Numerically computing the fine-scale concentration requires some care. From Eqs.~\eqref{eq::interp} and~\eqref{eq:c_fine}, we have
\begin{align}
	c(x,t) &= \sum_i \int\limits_0^t dt' \, \mathcal K_i(x,t') P_i(t-t'),
\end{align}
where the kernel $\mathcal K_i$ is defined by its Laplace transform,
\begin{align}
\tilde{\mathcal K}_i(x,\lambda) = \frac{ \lambda \tilde
  g_i(x-x_i,\lambda)}{1 - \tilde \psi_i(\lambda)}. 
\end{align}
We now write the fine-scale concentration as
\begin{align}
	c(x,t) = \sum_i \sum_{j=0}^{N-1} \int\limits_{t_j}^{t_{j+1}} dt' \, \mathcal K_i(x,t') P_i(t-t'),
\end{align}
where $t_j = j \Delta t$ and $N=t/\Delta t$.
The time discretization
$\Delta t$ is chosen on the order of the smallest characteristic transition time, so that the coarse-grained node occupation probabilities $P_i$ vary little within a time step. The memory kernels $\mathcal K_i$, on the other hand, may vary quickly at some positions. We thus employ the following approximation:
\begin{equation}
	c(x,t) \approx \sum_i \sum_{j=0}^{N-1} P_i(t-t_{j+1}) \int\limits_{t_j}^{t_{j+1}} dt' \, \mathcal K_i(x,t').
\end{equation}
The remaining integral can be computed using numerical Laplace inversion by considering that
\begin{equation}
	\mathcal{L}_t \left[\int\limits_0^t dt' \, \mathcal K_i(x,t') \right](\lambda) = \frac{\tilde{\mathcal K}_i(x,\lambda)}{\lambda},
\end{equation}
where $\mathcal{L}_t$ is the Laplace operator in $t$, $\mathcal{L}_t [f(t)](\lambda) = \tilde f(\lambda)$, and
\begin{align}
\notag
	\int\limits_{t_j}^{t_{j+1}} dt' \, \mathcal K_i(x,t') &= \int\limits_{0}^{t_{j+1}} dt' \, \mathcal K_i(x,t')\\
	&\quad- \int\limits_{0}^{t_j} dt' \, \mathcal K_i(x,t').
\end{align}

\section{Diffusion under trapping in power-law media}
\label{s::trapping}

We apply the TDRW method developed here to diffusion under a broad distribution of retardation coefficients $\theta(x)$ and heterogeneity length scales. The transport problem is described by the Langevin equation~\eqref{eq::ptrw}. We compare TDRW simulations to detailed PTRW simulations. In order to speed up the PTRW simulations, we perform a variable transform $t \rightarrow s$ with $dt = \theta[x(t)] ds$, so that Eq.~\eqref{eq::ptrw} reads
\begin{align}
\label{eq:var_change}
	d X(s) = \sqrt{2\kappa \, ds} \, \xi(s),&& d t(s) = \theta[X(s)] \, d s.
\end{align}
Note that this transformation renders the particle tracking equation a time domain random walk, because the time increment varies randomly. However, unlike the coarse-grained TDRW~\eqref{eq::tdrw}, the space steps are not synchronized with medium geometry and instead resolve the detailed particle motion.

We consider regions of constant retardation with lengths and retardation coefficient values distributed according to Pareto densities with infinite mean,
\begin{subequations}
\begin{align}
	\label{eq::length_dist}
	p_\ell(\ell) &= \ell_0 \left(\frac{\ell}{\ell_0}\right)^{-(1+\alpha)} \Theta(\ell-\ell_0), & 0<\alpha<1,\\
	\label{eq::retardation_dist}
	p_\theta(\theta) &= \theta_0 \left(\frac{\theta}{\theta_0}\right)^{-(1+\beta)} \Theta(\theta-\theta_0), & 0<\beta<1.
\end{align}
\end{subequations}
We take all segment lengths and retardation coefficients to be independent and identically distributed.
We set the minimum length $\ell_0=1$, the minimum retardation coefficient $\theta_0 = 1$, and the diffusion coefficient $\kappa = 1/2$, which is equivalent to normalizing lengths by $\ell_0$ and times by the smallest characteristic transition time $\tau_0 = \ell_0^2 \theta_0 / (2\kappa)$.

\begin{figure}[!bth]
\centering
\includegraphics[width=1\columnwidth]{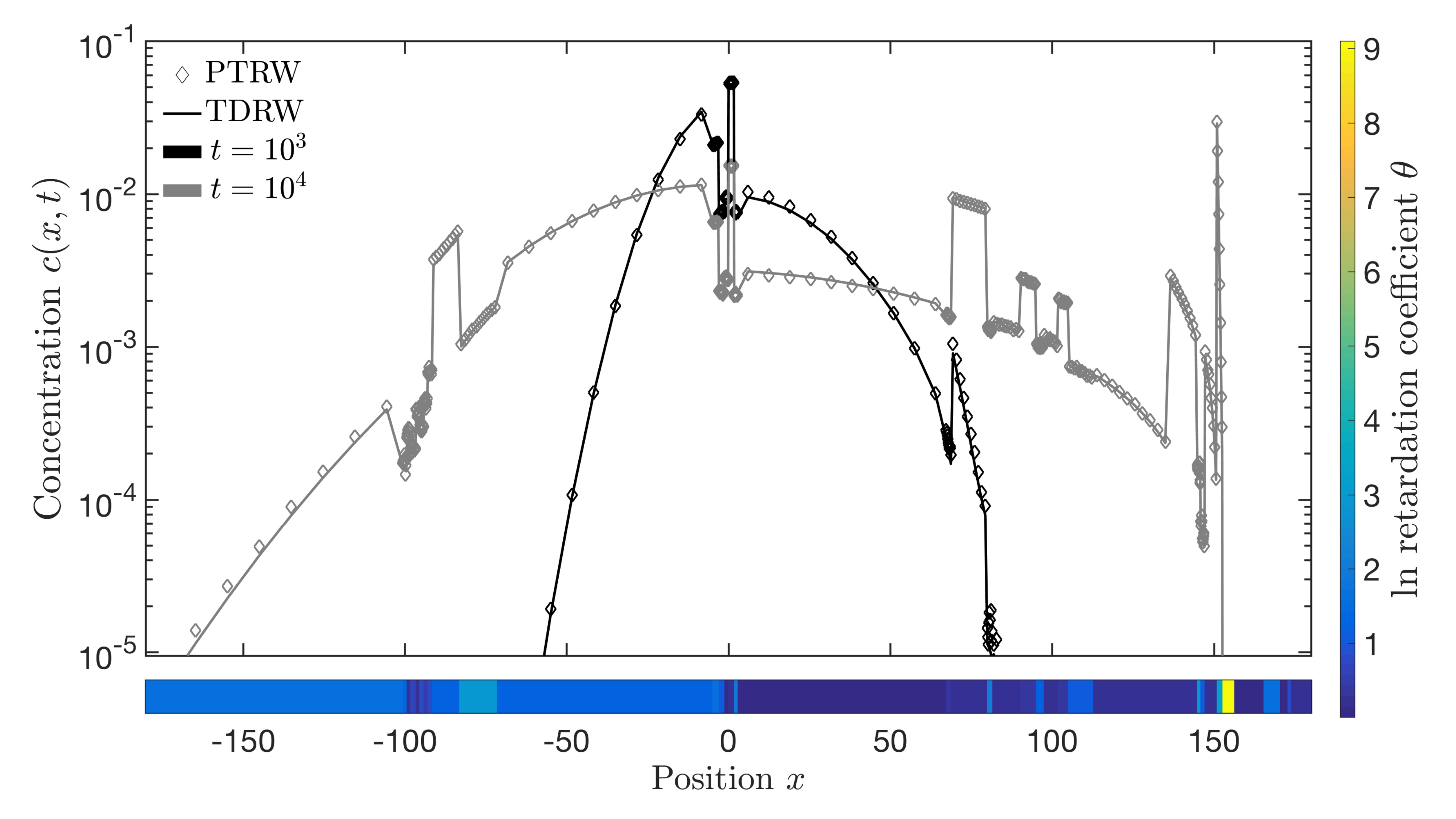}
\caption{Comparison of concentrations computed from PTRW and fine-scale TDRW simulations. We used $10^7$ particles for both approaches. We take infinite-mean length and retardation distributions with $\alpha = 0.7$ and $\beta = 0.8$. Results are for a single medium realization with segment lengths and retardation coefficients as shown.}
\label{f:comp}
\end{figure}

The PTRW simulations are discretized with $\Delta s = 2.5 \cdot 10^{-3}$, so that the characteristic jump size $\sqrt{2\kappa \Delta s} \ll \ell_0 = 1$. For the TDRW simulations, we solve for the node probability masses $P_i(t)$ using the Lagrangian equations~\eqref{eq::tdrw}, and record their values at intervals of $\Delta t = \tau_0 = 1$. For the transition times, we employ the truncated inverse Gaussian approximation~\eqref{eq::fpt_tig}. The reconstruction of the fine-scale concentration is described in the previous section.
Figure~\ref{f:comp} shows the comparison between concentration distributions obtained from PTRW simulations and the TDRW approach at two different times, along with the realization of the heterogeneous medium in which the transport is simulated. The reconstructed TDRW results are in excellent agreement with the PTRW data, and the TDRW simulations are much more efficient than the PTRW simulations, even using Eq.~\eqref{eq:var_change}. For example, for the concentration at $t=10^4$, the TDRW simulation with the fine-scale concentration reconstructed at $10$ positions for each segment between two nodes is about $5\e{2}$ times faster than the PTRW simulation. The coarse simulation, which renders the probability masses $P_i(t)$, is about $10^3$ times faster than the PTRW simulation.
 
In the following, we analyze the role of the reconstruction of the fine scale concentration in more detail by looking at the spatial moments of concentration. We show how ensemble-averaged (i.e., averaged over disorder realizations) moments may be estimated directly from the coarse-grained TDRW simulations based on~\eqref{eq::tdrw} and find that the coarse-grained description preserves the correct late-time scaling of the ensemble-averaged plume variance.

Details on the following calculations for the spatial mean and variance of concentration are given in~\ref{a::moments}. Denote the $k$th single-realization moment as a function of time $t$ for the coarse-grained TDRW by $m_{c,k}(t)$ and its fine-scale counterpart by $m_{f,k}(t)$,
\begin{align}
m_{c,k}(t) = \sum_i x_i^k P_i(t), &&m_{f,k}(t) = \int dx \, x^k c(x,t).
\end{align}
In Laplace space, we have
\begin{subequations}
\begin{align}
	\tilde m_{c,k}(\lambda) &= \sum_i x_i^k \tilde P_i(\lambda),\\
	\tilde m_{f,k}(\lambda) &= \lambda \sum_i \frac{\tilde P_i(\lambda)}{1-\tilde\psi_i(\lambda)} \int dx \, x^k \tilde g_i(x-x_i,\lambda),
\end{align}
where we have used Eq.~\eqref{eq::interp}. 
\end{subequations}
The mean of the fine-scale concentration is identically zero,
\begin{align}
m_{f,1}(t) \equiv 0.
\end{align}
The mean of the coarse-grained TDRW is obtained using Eq.~\eqref{eq::meq_waiting} as
\begin{align}
\notag
	\tilde m_{c,1}(\lambda) &= \sum_i \frac{\ell_{+,i}\ell_{-,i}}{\ell_{+,i}+\ell_{-,i}} \left[  \tilde \psi_{+,i}(\lambda) - \tilde \psi_{-,i}(\lambda) \right]\\
	&\quad\times\frac{\tilde P_i(\lambda)}{1-\tilde\psi_i(\lambda)}.
\end{align}
The coarse-grained TDRW produces a spurious term in the average position due to the difference between $\psi_+$ and $\psi_-$, the transition time PDFs given the transition direction. This is because, while each transition length $L_n$ has zero mean, the contribution to the mean of particles that have undergone different numbers of transitions may be nonzero when not considering in-cell particle transport.

For the second moments, we obtain
\begin{subequations}
\begin{align}
 \notag
	\tilde m_{c,2}(\lambda) &= \sum_i \bigg\{ l^2_{+,i} \tilde \phi_{+,i}(\lambda) + l^2_{-,i} \tilde \phi_{-,i}(\lambda)\\ \notag
		&\quad+2 x_i \left[ \ell_{+,i} \tilde \phi_{+,i}(\lambda) - \ell_{-,i} \tilde \phi_{-,i}(\lambda) \right] \bigg\}\\
		&\quad\times\frac{\tilde P_i(\lambda)}{1-\tilde\psi_i(\lambda)},\\
\notag
	\tilde m_{f,2}(\lambda) &= 2\lambda^{-1}\sum_i\sum_{\sigma\in\{+,-\}}\frac{\tilde P_i(\lambda)}{1-\tilde\psi_i(\lambda)}\\
	&\quad\times D_{\sigma,i}\left[\cosh(\ell_{\sigma,i}\sqrt\frac{\lambda}{D_{\sigma,i}})-1\right]\tilde\phi_{\sigma,i}(\lambda).
\end{align}
\end{subequations}

We see that the coarse-grained TDRW produces spurious contributions to the moments because in-cell concentration distributions are not taken into account. Nevertheless, we will now show how the coarse-grained TDRW simulations can be used to directly estimate the ensemble-averaged moments of the full process.
The spurious term for the coarse-grained average disappears in the ensemble average due to symmetry in $\ell_{\pm,i}$ and $D_{\pm,i}$. For the case of the second moment, the surviving correction under the ensemble average requires a more detailed analysis.

\subsection{Finite-mean transition time}

If the ensemble-averaged (denoted by an overline) mean transition time $\overline{\mu}$ is finite, we can Taylor-expand the second moments for $\lambda\ll 1/\overline{\mu}$ to find to leading order
\begin{align}
	\overline{\tilde m_{f,2}(\lambda)} \approx \overline{\tilde m_{c,2}(\lambda)} \approx \frac{2}{\lambda}\sum_i \overline{D_{\mathrm{eff},i} \tilde P_i(\lambda)},
\end{align}
so that at late times $t \gg \overline{\mu}$ the coarse-grained model gives the same results as the fine-scale version,
\begin{align}
\label{eq:var_finite_mean}
	\overline{m_{f,2}(t)} \approx \overline{m_{c,2}(t)} \approx 2 \int\limits_0^t \, dt' \, \overline{D_\mathrm{eff}(t')},
\end{align}
where $D_\mathrm{eff}(t) = \sum_i D_{\mathrm{eff},i} P_i(t)$ is the average effective dispersion sampled by the concentration plume in a single medium realization.

\subsection{Infinite-mean transition time}

In order to study what happens when the segment lengths have infinite variance, in which case the previous expansion breaks down,
we take segment lengths to follow a Pareto density with infinite mean as above, see Eq.~\eqref{eq::length_dist}. The correction may be written as
\begin{align}
\notag
	\overline{\delta\tilde m_{2}(\lambda)} &= \overline{\tilde m_{f,2}(\lambda)} - \overline{\tilde m_{c,2}(\lambda)},\\
		&= \overline{\sum_i \frac{\int dx\, x^2 \tilde g_i(x,\lambda)}{\int dx\, \tilde g_i(x,\lambda)} \tilde P_i(\lambda)}.
\end{align}
It represents the contribution to the variance of particles diffusing within their current unit cell before completing the current transition. 

Computing this correction directly is difficult, and we estimate it as follows. Because particles are likely to spend longer times in larger segments, and in turn these contribute the most to the variance correction, we will estimate the correction by the square of the largest segment length seen by a particle up to a given time. We estimate the size of the largest out of $n$ segments, $\ell_\mathrm{max,n}$, through~\cite{bouchaud1990anomalous}
\begin{equation}
	\int\limits_{\ell_{\mathrm{max},n}}^\infty d\ell \, p_\ell(\ell) \approx \frac{1}{n},
\end{equation}
which says that the probability of a value larger than $ \ell_{\mathrm{max},n}$ is approximately $1/n$. This gives $\ell_{\mathrm{max},n} \approx \ell_0 n^\frac{1}{\alpha}$. Adapting the result for waiting times in~\cite{benson2007recurrence}, the Laplace transform over $x$ of the probability of having exactly $n$ full segments contained within a region of length $x$ is given by
\begin{equation}
	\tilde q_n(k) = \frac{1-\tilde p_\ell(k)}{k}[\tilde p_\ell(k)]^n.
\end{equation}
For small $k\ll 1/\ell_0$ (large $x \gg \ell_0$), we have $\tilde p_\ell(k) \approx 1 - |\Gamma(-\alpha)| (\ell_0 k)^\alpha$, where $\Gamma(\cdot)$ is the gamma function.
We thus find for the average number of segments
\begin{equation}
	\overline{\tilde N(k)} = \sum_{n\geqslant1} n \tilde q_n(k) \approx \frac{\ell_0(\ell_0k)^{1-\alpha}}{\alpha |\Gamma(-\alpha)|}.
\end{equation}
Inverting the Laplace transform,
\begin{equation}
	\overline{N(x)} \approx \frac{\sin(\pi\alpha)}{\pi\alpha} \left(\frac{x}{\ell_0}\right)^\alpha,
\end{equation}
where we have used Euler's reflection formula, $\Gamma(z)\Gamma(1-z) = \pi z / \sin(\pi z)$, with $z=-\alpha$.
This allows us to estimate the maximum segment length in a region of length $x$, $\ell_\mathrm{max}(x) = \ell_{\mathrm{max},\overline{N(x)}}$, giving
\begin{equation}
	\ell^2_\mathrm{max}(x) \approx \left(\frac{\sin(\pi\alpha)}{\pi\alpha}\right)^{2/\alpha} x^2.
\end{equation}
The typical distance $x$ covered by a particle by time $t$ is on the order of the standard deviation $\overline{m_{c,2}(t)}^{1/2}$. Thus, setting $x(t) = a\overline{m_{c,2}(t)}^{1/2}$, we obtain 
\begin{equation}
\label{eq::var_interp}
	\overline{\delta \tilde m_2(t)}  \approx a^2\left(\frac{\sin(\pi\alpha)}{\pi\alpha}\right)^{2/\alpha} \overline{m_{c,2}(t)},
\end{equation}
from which we find that $\overline{m_{f,2}(t)} \propto \overline{m_{c,2}(t)}$. This argument predicts the late-time scaling behavior of the coarse-grained model to be the same as that of its fine-scale counterpart. The correction approaches zero as $\alpha$ approaches unity, which suggests that when the length average is finite the coarse-grained approach captures the plume variance correctly even if the variance is infinite. Indeed, the same argument as above for $1<\alpha<2$, for which $\tilde p_\ell(k) \approx 1 - \alpha \ell_0 k/(\alpha-1)$, yields to subleading correction.

\begin{figure*}[!htb]
\centering
\includegraphics[width=1\textwidth]{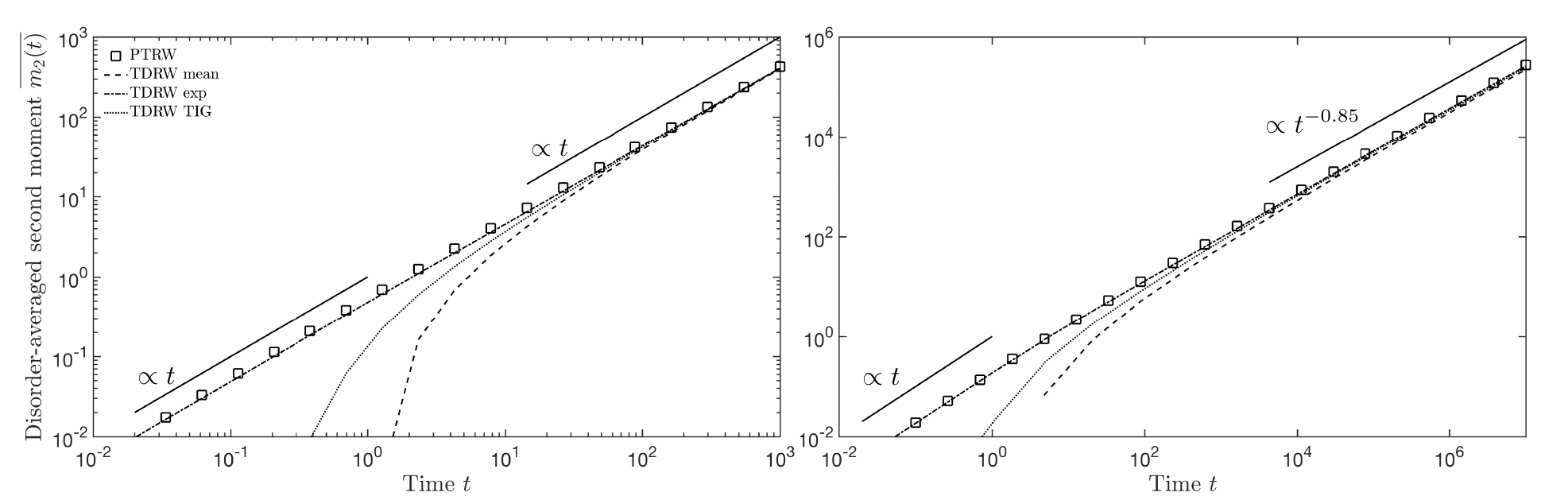}
\caption{Comparison between PTRW and TDRW computations of the ensemble-averaged second moment of concentration, using the different approximations for the TDRW transition times discussed in the text. A sufficient number of segments ($10^3$) was used so that no particles reached the domain boundaries. For the PTRW (respectively TDRW) simulations, we used $10^2$ ($10^3$) particles in $10^3$ ($10^4$) medium realizations. Left: $\alpha=1.7$ and $\beta=1.5$ (finite-mean transition time), with TDRW results computed directly from the coarse-grained concentration. Right: $\alpha=0.6$ and $\beta=0.4$ (infinite-mean transition time), with coarse-grained TDRW results rescaled according to Eq.~\eqref{eq::var_interp} with $a=2.5$.}
\label{fi::comparison}
\end{figure*}

We now compare our results for the ensemble-averaged plume variance against numerical simulations. We compute the variance directly from the coarse-grained model, and include the correction~\eqref{eq::var_interp} for infinite-mean length, see Fig.~\ref{fi::comparison}. We find that $a = 2.5$ provides good agreement with simulations when the mean length is infinite.
%
%
%
We use the same setup as in Section~\ref{s::interp}. For the transition times, we employ the approximations introduced in Section~\ref{s::tdrw_fpt}, Eqs.~\eqref{eq::fpt_mean}--\eqref{eq::fpt_tig}.
We see that all these approximations yield similar late-time behaviors for $\overline{m_2(t)}$. Although the truncated inverse Gaussian approximation is better suited for reconstructing the fine-scale concentration in single medium realizations, the exponential approximation provides the best results for the ensemble-averaged moments at early times. This is because it effectively reproduces the diffusive scaling regime that corresponds to transport within a single unit cell at early times~\cite{russian2017self}.

\section{Summary and Conclusions}
We have derived a TDRW model for transport in media characterized by broad distributions of heterogeneity length scales. The transition lengths in this approach are determined by the medium geometry as spatial regions in which the transport properties are constant, and the transition times vary according to the length scales and transport properties. This coarse-grained TDRW is efficient, but it does not resolve the concentration variability below the discretization scale. Thus, we have developed a reconstruction procedure to determine the fine-scale concentration distribution using the propagator of the detailed transport problem within a unit cell, which constitutes a composite medium with two segments of different lengths and transport properties. Our formulation provides an efficient approach for modeling advective--dispersive transport in media with broadly distributed heterogeneity length scales and transport properties.

In the absence of advection, the transition probabilities in the retardation problem depend only on the segment length, with a higher probability to make a transition in the direction of the shorter segment. For the dispersion problem, the transition probability depends both on the segment lengths and dispersion properties. We explicitly determine the joint distribution of spatial and temporal transitions and show that the transition time distribution is well approximated by an inverse Gaussian distribution parameterized by the mean and variance of the transition times, which are both known analytically in terms of segment lengths and transport properties.

We illustrate the TDRW approach for diffusion under random retardation in a power-law medium, i.e., a medium characterized by heavy-tailed distributions of length scales and transport properties. The reconstructed fine-scale concentration using the truncated inverse Gaussian approximation for transition times is in excellent agreement with detailed simulations. In the ensemble sense, we find anomalous dispersive behavior for broadly distributed (infinite-mean) transition times, with sublinear scaling of the displacement variance. We demonstrate that both the coarse- and fine-scale TDRW approaches yield accurate predictions of the scaling of the displacement variance. In particular, this means that the coarse-scale TDRW provides a fast and efficient technique for probing the global transport dynamics. Analytically predicting the late-time scaling behavior of the plume moments from the coupled TDRW proposed here remains an open problem to be addressed in further work.
We also expect a generalization to unit cells in multiple dimensions and with more than two neighbors per node to be a promising modeling approach for transport in complex, multidimensional heterogeneous media and networks.

\section*{Acknowledgments}
The authors acknowledge the support of the European Research Council (ERC) through the project MHetScale (617511).

\appendix

\section{Unit cell}
\label{a::unit_cell}

We describe here in detail the steps for the derivation of the concentration propagator and arrival time distributions in the unit cell for the trapping problem. The steps are completely analogous for the dispersion problem.

We derive explicit solutions in Laplace space. Taking the Laplace transform over the time variable (denoted by a tilde) of Eq.~\eqref{eq::trapping} with a unit-mass pulse at the origin as the initial condition gives the governing equation for the concentration propagator $g$,
\begin{equation}
\label{eq::trapping_laplace}
	\lambda \tilde g(x,\lambda) + \frac{\partial}{\partial x}[v(x) \tilde g(x,\lambda)] - \frac{\partial^2}{\partial x^2}[D(x) \tilde g(x,\lambda)] = \delta(x).
\end{equation}
This is an equation for a Green function, which is to be solved with absorbing boundary conditions on the unit cell as discussed in the main text. The solutions are of the form $g(x,t) = g_+(x,t)\Theta(x) + g_-(x,t) \Theta(-x)$, where in Laplace space
\begin{align}
\notag
	\tilde g_\pm(x,\lambda) &= A_\pm(\lambda) e^{\frac{v_\pm}{2D_\pm}x[1-\alpha_\pm(\lambda)]}\\
	&\quad+ B_\pm(\lambda) e^{\frac{v_\pm}{2D_\pm}x[1+\alpha_\pm(\lambda)]},
\end{align}
where $\alpha_\pm(\lambda) = \operatorname{sgn}(v_\pm)\sqrt{1+4D_\pm \lambda/v_\pm^2}$, with $\operatorname{sgn}(v)$ denoting the sign of $v$.
Using the boundary conditions, we obtain
\begin{align}
\notag
	\tilde g_\pm(x,\lambda) &= A_\pm(\lambda)e^{\frac{v_\pm}{2D_\pm}x}\bigg( e^{-\frac{v_\pm}{2D_\pm}x \alpha_\pm(\lambda)}\\
	&\quad-e^{\frac{v_\pm}{2D_\pm}(x\mp2\ell_\pm) \alpha_\pm(\lambda)} \bigg).
\end{align}
Using the trapping continuity condition $D_- g_- = D_+ g_+$ at $x=0$, as discussed in the main text, we obtain
\begin{subequations}
\begin{align}
 \notag
	\tilde g_-(x,\lambda) &= A_-(\lambda)e^{\frac{v_-}{2D_-}x}\bigg( e^{-\frac{v_-}{2D_-}x \alpha_-(\lambda)}\\
	&\quad- e^{\frac{v_-}{2D_-}(x+2\ell_-) \alpha_-(\lambda)} \bigg),\\ \notag
	\tilde g_+(x,\lambda) &= A_-(\lambda)\frac{D_-}{D_+}\frac{1-e^{\frac{\ell_-v_-}{D_-}\alpha_-(\lambda)}}{1-e^{-\frac{\ell_+v_+}{D_+}\alpha_+(\lambda)}}e^{\frac{v_+}{2D_+}x}\\ \notag
	&\quad\times \bigg( e^{-\frac{v_+}{2D_+}x \alpha_+(\lambda)}\\
	&\quad- e^{\frac{v_+}{2D_+}(x-2\ell_+) \alpha_+(\lambda)} \bigg).
\end{align}
\end{subequations}

Now we integrate Eq.~\eqref{eq::trapping_laplace} over $x$, which gives
\begin{align}
\notag
	\lambda \tilde M(\lambda) &= 1 + D_+ \frac{\partial \tilde g_+(x,\lambda)}{\partial x}\bigg|_{x=\ell_+}\\ \notag
	&\quad- D_- \frac{\partial \tilde g_-(x,\lambda)}{\partial x}\bigg|_{x=-\ell_-},\\ \notag
	&=  1 + A_-(\lambda)\\ \notag
	&\quad\times\Bigg[ v_- \alpha_-(\lambda) e^{-\frac{\ell_- v_-}{2D_-}[1-\alpha_-(\lambda)]}\\ \notag
	&\quad- \frac{D_-}{D_+} v_+\alpha_+(\lambda) e^{\frac{\ell_+ v_+}{2D_+}[1-\alpha_+(\lambda)]}\\
	&\quad\times\frac{1-e^{\frac{\ell_-v_-}{D_-}\alpha_-(\lambda)}}{1-e^{-\frac{\ell_+v_+}{D_+}\alpha_+(\lambda)}} \Bigg],
\end{align}
where $M(t)$ is the total mass in the unit cell at time $t$. On the other hand, integrating Eq.~\eqref{eq::sol_piecewise} and multiplying by $\lambda$ we get
\begin{align}
\notag
	\lambda \tilde M(\lambda) &= \frac{A_-(\lambda)}{2}\Bigg\{ v_+ \frac{D_-}{D_+} \frac{1-e^{\frac{\ell_-v_-}{D_-}\alpha_-(\lambda)}}{1-e^{-\frac{\ell_+v_+}{D_+}\alpha_+(\lambda)}}\\ \notag
	&\quad\times\Bigg[ 1 - e^{-\frac{\ell_+ v_+}{D_+}\alpha_+(\lambda)}+ \alpha_+(\lambda)\\ \notag
	&\quad\times\bigg( 1 - 2 e^{\frac{\ell_+ v_+}{2D_+}[1-\alpha_+(\lambda)]} + e^{-\frac{\ell_+ v_+}{D_+}} \bigg) \Bigg]\\ \notag
	&\quad- v_- \Bigg[ 1 - e^{\frac{\ell_- v_-}{D_-}\alpha_-(\lambda)}+ \alpha_-(\lambda)\\
	&\quad\times\bigg( 1 - 2 e^{-\frac{\ell_- v_-}{2D_-}[1-\alpha_-(\lambda)]} + e^{\frac{\ell_- v_-}{D_-}} \bigg) \Bigg] \Bigg\}.
\end{align}
Eliminating $\tilde M(\lambda)$ and solving for $A_-(\lambda)$ using these two equations we subsequently obtain the results in Table~\ref{t::unit_cell} through straightforward manipulations. We omit the calculations for regular dispersion for brevity. The approach is analogous, but enforcing continuity of the concentration at $x=0$ as discussed in the main text.

\section{Moments}
\label{a::moments}

Here we present some details on the calculations of the moments for the diffusive trapping problem discussed in Section~\ref{s::trapping}.

For the coarse-grained average, using Eq.~\eqref{eq::meq_waiting}, renaming indices under the sums and noting that $x_i \pm \ell_{\pm,i} = x_{i\pm1}$, we have
\begin{align}
\notag
	\tilde m_{c,1}(\lambda) &= \sum_i x_i \tilde P_i(\lambda),\\ \notag
		&= \sum_i \left[ \ell_{+,i} \tilde \phi_{+,i}(\lambda) - \ell_{-,i} \tilde \phi_{-,i}(\lambda) \right] \frac{\tilde P_i(\lambda)}{1-\tilde\psi_i(\lambda)} ,\\ \notag
		&= \sum_i \frac{\ell_{+,i}\ell_{-,i}}{\ell_{+,i}+\ell_{-,i}} \left[  \tilde \psi_{+,i}(\lambda) - \tilde \psi_{-,i}(\lambda) \right]\\
		&\quad\times\frac{\tilde P_i(\lambda)}{1-\tilde\psi_i(\lambda)}.
\end{align}
Similarly, for the coarse-grained second moment we have
\begin{align}
\notag
	\tilde m_{c,2}(\lambda) &= \sum_i \bigg\{ l^2_{+,i} \tilde \phi_{+,i}(\lambda) + l^2_{-,i} \tilde \phi_{-,i}(\lambda)\\ \notag
		&\quad+2 x_i \left[ \ell_{+,i} \tilde \phi_{+,i}(\lambda) - \ell_{-,i} \tilde \phi_{-,i}(\lambda) \right] \bigg\}\\
		&\quad\times\frac{\tilde P_i(\lambda)}{1-\tilde\psi_i(\lambda)}.
\end{align}

For the fine-scale moments, we make use of the following results:
\begin{subequations}
\begin{align}
	\int\limits_{\Omega_i} dx\, \tilde g_i(x,\lambda) &= \frac{1-\tilde\psi_i(\lambda)}{\lambda},\\
	\int\limits_{\Omega_i} dx\, x \tilde g_i(x,\lambda) &= \frac{\ell_{-,i}\tilde \phi_{-,i}(\lambda) - \ell_{+,i}\tilde \phi_{+,i}(\lambda)}{\lambda},\\ \notag
	\int\limits_{\Omega_i} dx\, x^2 \tilde g_i(x,\lambda) &= 2\lambda^{-2}\sum_i\sum_{\sigma\in\{+,-\}}\\ \notag
	&\quad\bigg[D_{\sigma,i}\left(\cosh(\ell_{\sigma,i}\sqrt\frac{\lambda}{D_{\sigma,i}})-1\right)\\
	&\quad-\ell_{\sigma,i}^2\lambda\bigg]\tilde\phi_{\sigma,i}(\lambda),
\end{align}
\end{subequations}
which can be obtained from the explicit expressions for the unit-cell propagator discussed in Section~\ref{s::fpt_unit_cell} and presented in Table~\ref{t::unit_cell}. From Eq.~\eqref{eq::interp}, the average is identically zero,
\begin{align}
\notag
	\tilde m_{f,1}(\lambda) &= \int dx \, x \sum_i \tilde R_i(\lambda) \tilde g_i(x-x_i,\lambda),\\ \notag
		&\quad+ \sum_i \tilde R_i(\lambda) \int\limits_{\Omega_i} dx \, x \tilde g_i(x,\lambda),\\
		&= \tilde m_{c,1}(\lambda) - \tilde m_{c,1}(\lambda) \equiv 0.
\end{align}
For the second moment,
\begin{align}
\notag
	\tilde m_{f,2}(\lambda) &= \sum_i \tilde R_i(\lambda) \int\limits_{\Omega_i} dx \, (x+x_i)^2 \tilde g_i(x,\lambda),\\ \notag
	&= 2\lambda^{-1}\sum_i\sum_{\sigma\in\{+,-\}}\frac{\tilde P_i(\lambda)}{1-\tilde\psi_i(\lambda)}\\
	&\quad\times D_{\sigma,i}\left[\cosh(\ell_{\sigma,i}\sqrt\frac{\lambda}{D_{\sigma,i}})-1\right]\tilde\phi_{\sigma,i}(\lambda).
\end{align}

Taking into account the previous derivations, and noting that some terms disappear under the ensemble average due to symmetry, it is straightforward to show that the ensemble-averaged difference $\overline{\delta m_{2}(t)} = \overline{m_{f,2}(t)} - \overline{m_{c,2}(t)}$ between the fine-scale and coarse-grained second moments has Laplace transform

\begin{align}
\notag
	\overline{\delta\tilde m_2(\lambda)}
		&= \lambda\overline{\sum_i \frac{\int dx\, x^2 \tilde g_i(x,\lambda)}{1-\tilde\psi_i(\lambda)} \tilde P_i(\lambda)},\\
		&= \overline{\sum_i \frac{\int dx \, x^2 g_i(x,\lambda)}{\int dx\, \tilde g_i(x,\lambda)} \tilde P_i(\lambda)},
\end{align}
where in the last equality we have used the Laplace transform of Eq.~\eqref{eq:cpsi}.

When the ensemble-averaged mean transition time $\overline{\mu}$ is finite, the quantities $l_{\pm,i}^2/D_{\pm,i}$ also have a finite ensemble average. For small $\lambda \ll 1/\overline{\mu}$, using Taylor expansions around $\lambda = 0$, we approximate
\begin{align}
\notag
	\tilde\psi_i(\lambda) &\approx 1 - \mu_i\lambda,\\\notag
	\tilde\phi_{\pm,i}(\lambda) &\approx p_\pm(1 - \mu_{\pm,i} \lambda),\\ \notag
	\cosh\left(\ell_{\pm,i}\sqrt\frac{\lambda}{D_{\pm,i}}\right) &\approx 1 + \frac{\ell^2_{\pm,i}\lambda}{2D_{\pm,i}},
\end{align}
and find that the ensemble-averaged coarse-grained and fine-scale second moments agree to leading order and are given by
\begin{align}
	\overline{\tilde m_{f,2}(\lambda)} \approx \overline{\tilde m_{c,2}(\lambda)} \approx 2\lambda^{-1}\sum_i \overline{D_{\mathrm{eff},i} \tilde P_i(\lambda)}.
\end{align}







\end{document}